# Devito: automated fast finite difference computation

Navjot Kukreja*, Mathias Louboutin†, Felippe Vieira*, Fabio Luporini‡, Michael Lange§, Gerard Gorman§

*SENAI CIMATEC, Salvador, Brazil Emails: navjotk@gmail.com, felippe.vieira@fieb.org.br
†Seismic Lab. for Imaging and Modeling, University of British Columbia Email: mloubout@eos.ubc.ca
‡Department of Computing, Imperial College London Email: f.luporini12@imperial.ac.uk
§Department of Earth Science and Engineering, Imperial College London Emails: {michael.lange, g.gorman}@imperial.ac.uk

*Abstract*—Domain specific languages have successfully been used in a variety of fields to cleanly express scientific problems as well as to simplify implementation and performance optimization on different computer architectures. Although a large number of stencil languages are available, finite difference domain specific languages have proved challenging to design because most practical use cases require additional features that fall outside the finite difference abstraction. Inspired by the complexity of real-world seismic imaging problems, we introduce Devito, a domain specific language in which high level equations are expressed using symbolic expressions from the *SymPy* package. Complex equations are automatically manipulated, optimized, and translated into highly optimized C code that aims to perform comparably or better than hand-tuned code. All this is transparent to users, who only see concise symbolic mathematical expressions.

## I. INTRODUCTION

It is well-known that the complexity of computer architectures tends to increase with their power. As a result, the software needed to obtain the desired performance has become more sophisticated. Code written for conventional CPUs is not the same as code written for accelerator platforms such as the Intel® Xeon Phi™[12] or GPUs [28]. In the past, new programming paradigms have been introduced to raise the level of abstraction, with the aim of writing performance portable software.

While the search for a generic abstraction for high performance on parallel architectures continues, domain specific languages (DSL) have proven particularly effective for the rapid development of efficient, maintainable, and portable scientific simulations. The main idea of this approach is to decouple the problem specification from its low level implementation through a stack of abstraction layers. This creates a separation of concerns between domain scientists and compiler/architecture specialists, which has a direct payoff in productivity.

In this paper we present Devito, a new finite difference (FD) DSL for seismic inversion problems that leverages the powerful *SymPy* Python package to define complex mathematical operators and auto-generates low-level optimized thread-parallel C code for multiple target architectures. Devito provides multiple API layers to allow users to fully leverage "pure" symbolic expressions and optimization, as well as a low-level mechanism to enable non-standard features. After providing a brief overview of related DSLs, we describe the multi-level DSL itself along with its key features before demonstrating the definition of realistic seismic inversion operators.

## II. RELATED WORK

Interest in building DSLs for solving partial differential equations (PDE) is not new with notable work dating as far back as 1970 [5]. By trying to be too general across various types of differential equations, this was probably not able to take advantage of domain specific optimizations. DEQSOL [24] was an early attempt (1985) at providing a language for a high-level mathematical description which would generate highly vectorizable Fortran code meant to solve a PDE. Alpal [7] used symbolic techniques to manipulate the mathematical equation, derive the finite difference formulation, optimize it using common sub-expression elimination (CSE) and finally generate Fortran/C code. Ctadel [26] in 1996 was a complete DSL for explicit finite difference schemes that automatically generated code in Fortran/C for distributed memory parallel systems. Like Alpal, Ctadel featured CSE, a technique that we also discuss in later sections of this paper. Looking at these past works, it seems that it might not be possible to have a single FD DSL that works for every possible application domain. This is because each problem domain requires certain low-level features that are not related to finite difference. Any rigid FD DSL would then face the problem of implementation of these low-level features. Recent advances in computing speed make Just-in-Time (JIT) code generation and compilation a viable approach that mitigates this risk to some extent. The approach then is to target a smaller domain of problems for the DSL - at least initially, to be able to solve real problems in that domain before moving on to other domains.

Currently, two prominent examples of the DSL paradigm are the FEniCS [14] and Firedrake [20] projects, which allow the expression of finite element methods through a mathematical syntax embedded in Python. One of the fundamental layers in Firedrake is PyOP2 [16], which exposes a language for the execution of numerical kernels over unstructured grids. In particular, PyOP2 generates C code, which is eventually translated into machine code by the host compiler.

The Simflowny project [1] provides an end-to-end framework for building models of physical systems using PDEs which employs code generation for high performance. Al-though this is targeted towards end-user physicists who will not be writing nearly any code at all.

The field of domain specific languages and libraries for methods based upon structured grids, or stencils, is wide and quite heterogeneous. SBLOCK [4] is a Python framework to express complex stencils arising in real-world computations. Low-level code for multi-node architectures is automatically generated from the Python specification. The Mint framework, based on pragma directives for GPU execution, has been used to accelerate a 3D earthquake simulation [25]. The OPS [21] domain specific active library(i.e., looks like as a traditional library, but code generation is employed) provides a high-level abstraction similar to that of PyOP2, although it specifically targets multi-block structured mesh computations. Just like Devito, OPS targets real-world applications. Other projects use code generation and auto-tuning to increase the performance of the generated code [11, 29, 8, 6, 10, 17], although it is not clear to what extent they might be used for production code.

Polyhedral compilers have become increasingly popular over the last decade to automatically optimize stencil codes. These tools attempt to optimize so called affine loop nests (i.e., all bounds and array indices must be affine expressions in the enclosing loop indices), typically for parallelism and data locality. PLUTO [3] is probably the most famous and commonly used example of such compilers.

Finally, the Halide project (domain specific language, compiler, and auto-tuning system), despite originally designed for image processing pipelines, now provides mechanisms and optimizations that are of potential interest for generic stencil computations [19].

## III. RAISING THE LEVEL OF ABSTRACTION

When designing DSL-based scientific tools the choice of abstraction layers and interfaces is crucial, since too narrow abstractions often hamper future development once the original features are exhausted, while all-encompassing abstractions often become too generic, complicated and hard to maintain. For this reason, Devito relies on *SymPy* as its symbolic top-level language to express equations, as well as its internal representation of stencil expressions. The original expression is modified symbolically several times to progressively resemble C code, thus increasing the stencil's complexity. Through this multi-stage process additional user interfaces at lower levels are created that may be used for the addition of non-standard model features, and future compatibility with external tools is simplified.

As discussed in the later sections, the need for a full-scale symbolic top-level language becomes evident when complex mathematical models are considered. Some stencils are inherently difficult to express as well as to optimize – a single expression may involve hundreds of operations.

## IV. DEVITO OVERVIEW

One of the guiding design concepts behind Devito is the *Principle of Graceful Degradation* that suggests that users should be able to circumvent limits to the top layer abstraction by utilising a lower-level API that might be more laborious but less restrictive. In the context of seismic imaging, a major restriction that other FD DSLs discussed before have is that they do not have support for sparse-point interpolation, an important requirement for source insertion and receiver sampling. This is one area where the lower-level API of Devito was used to create a complete representation of the real problem. A second cornerstone in the design of Devito is the extensive use of *SymPy* to express, manipulate and evolve mathematical expressions from high-level equations to low-level stencil expressions that easily translate into C code. Ultimately, since Devito operators consist of purely symbolic expressions, low-level implementation choices that aid performance are left to Devito, which enables streamlined code generation bespoke to the target platform.

### A. Symbolic stencil equations

The *SymPy* Python package provides the basic symbol objects to represent complete equations in symbolic form. Devito utilises these expressions as the basis for generating optimized finite difference stencil operators by providing two types of objects:

**Symbolic Data** Objects that associate grid data with *SymPy* symbols to form symbolic expressions.
**Operator** Objects that generate optimized C code to apply a given stencil expression to the associated data.

Devito's symbolic data objects decorate user data with the symbolic behaviour of sympy.Function objects and encapsulate time-varying functions as well as field data constant in time. These objects inherit their symbolic behaviour directly from *SymPy* data objects and can be used directly within *SymPy* expressions with appropriate behaviour. For example, the following code shows a simple equation that updates one spatial grid with the values of another:

```
from Devito import DenseData, Operator
from sympy import Eq

f = DenseData(name='f', shape=(10, 10))
g = DenseData(name='g', shape=(10, 10))
h = DenseData(name='h', shape=(10, 10))

eqn = Eq(f, h + 2 * g)
op = Operator(stencils=eqn, timesteps=1)
```

In this code snippet the data objects f and g represent spatial functions and act as *SymPy* symbols f(x, y) and g(x, y), and the sympy.Equation object eqn is equivalent to the mathematical expression:

$$f(x,y) = h(x,y) + 2g(x,y).$$

Devito decorates its custom symbolic objects by allocating memory to store associated data along with the symbolic form of the variable. The encapsulated data is made accessible to users as a wrapped *NumPy* array object, allowing the initial values of the function to be set either via direct array access or by defining a custom initializer function that can be passed into the object constructor.

Using the above example, the user may set the initial values of the spatial fields via `f.data[:] = 5` and `g.data[:] = 3`. The `devito.Operator` object `op` now generates the corresponding C code for the equation `eqn` and when applied results in the function `f` having the final value of 11.

*1) Derivative expressions:* In addition to associated memory, Devito data objects also provide shorthand notation for common finite difference formulations. These consist primarily of automatically expanded symbolic expressions for first, second and cross derivatives in the time and space dimensions, where the order of the discretization is defined on the symbolic data object. For example defining the function $f$ as `f = DenseData('f', shape(10, 10), space_order=2)` allows us to take the second derivative of $f$ in dimension $x$ as:

```
In [1]: from devito import DenseData

In [2]: f = DenseData(name='f',
   shape=(10, 10), space_order=2)

In [3]: f.dx2
Out[3]: -2*f(x, y)/h**2 + f(-h + x,
   y)/h**2 + f(h + x, y)/h**2
```

The symbol `h` has been inserted to represent the grid spacing in $x$. It is important to note here that this API allows the user to change the spatial discretization of the problem by simply changing one single constructor parameter for the `DenseData` object `f`.

Another utility property of Devito's symbolic data objects is the `f.laplace` operator that expands to `(f.dx2 + f.dy2)` for two-dimensional problems and `(f.dx2 + f.dy2 + f.dz)` for three-dimensional ones. This allows the expression of PDEs, such as the acoustic wave equation, to be defined concisely:

```
u = TimeData(name='u', space_order=6, ...)
m = DenseData(name='m', space_order=6,
   ...)

wave_eqn = Eq(m * u.dt2, u.laplace)
```

*2) Custom stencil API:* While PDEs on field variables are naturally expressible as *SymPy* equations with Devito stencil derivatives, other core features required for seismic inversion simulations, such as sparse point interpolation, cannot be expressed as a single equation on `sympy.Function` objects. For this reason Devito offers a secondary API that allows custom expressions where variable objects are of type `sympy.Indexed`, to closely mirror C-style variable indexing. It is important to note that the finite difference spacing variable used for function indexing is replaced with discrete grid indices in this notation, as shown below.

```
# High-level expression equivalent to
   f.dx2
(-2*f(x, y) + f(x - h, y) + f(x + h, y))
   / h**2

# Low-level expression with explicit
   indexing
(-2*f[x, y] + f[x - 1, y] + f[x + 1, y])
   / h**2
```

A notable detail about the multi-layered expression API in Devito is that the low-level expanded expression is always generated within Devito as part of the code generation process. This is encapsulated in the `devito.Propagator` class that encapsulates the low-level code generation process, while most high-level symbolic operations are performed by `devito.Operator` objects. The combination of these two descriptor objects, as well as a set of compiler presets for a range of common architectures creates an abstraction cascade that incrementally lowers a high-level problem definition into native executable C code, as shown in Figure 1. It is important to note here that custom formulations can be injected at either level, allowing the user to combine high-level derivative expressions with custom *SymPy* expressions that may include explicit data indexing. This combination of APIs provides graceful abstraction degradation and enables rapid prototyping for future features.

### B. Automated performance optimization

To generate a performance optimized FD kernel, Devito uses properties of the target hardware and the dataset at the time of code generation. Looking at Listing 3, it can be seen that the code generated by Devito includes hard-coded data-specific values like the loop limits. Since the data specific properties are known at the time of code generation, putting this in the code helps the compiler in further optimising the loops since the limits are known.

*1) Vectorization:* The code generated by Devito uses parallelism at multiple levels to fully utilise modern multicore CPU architectures. At the data level the generated code contains compiler specific hints that ensure vectorization on different compilers. Since the allocation of memory is also controlled by Devito, we ensure that the memory is allocated on aligned boundaries, depending on the page size of the target platform. The aligned memory allocation combined with the compiler hints ensure that the performance of this autogenerated code is comparable to hand-optimized code [2].

*2) Parallelization:* The next level of parallelism is to use all the available cores of the target system. For this, the code can be generated with OpenMP `pragmas` inserted at appropriate locations to ensure parallelism.

After using all available cores, further optimizations are required for multi-socket or Non-Uniform Memory Access (NUMA) systems. Here again, having control over the data structures and their memory allocation makes this optimization easy to implement. The generated code in the framework includes an initialization step which exploits the *first-touch* [23] policy to make sure the memory is allocated close to the physical location of the corresponding thread which will be accessing it.

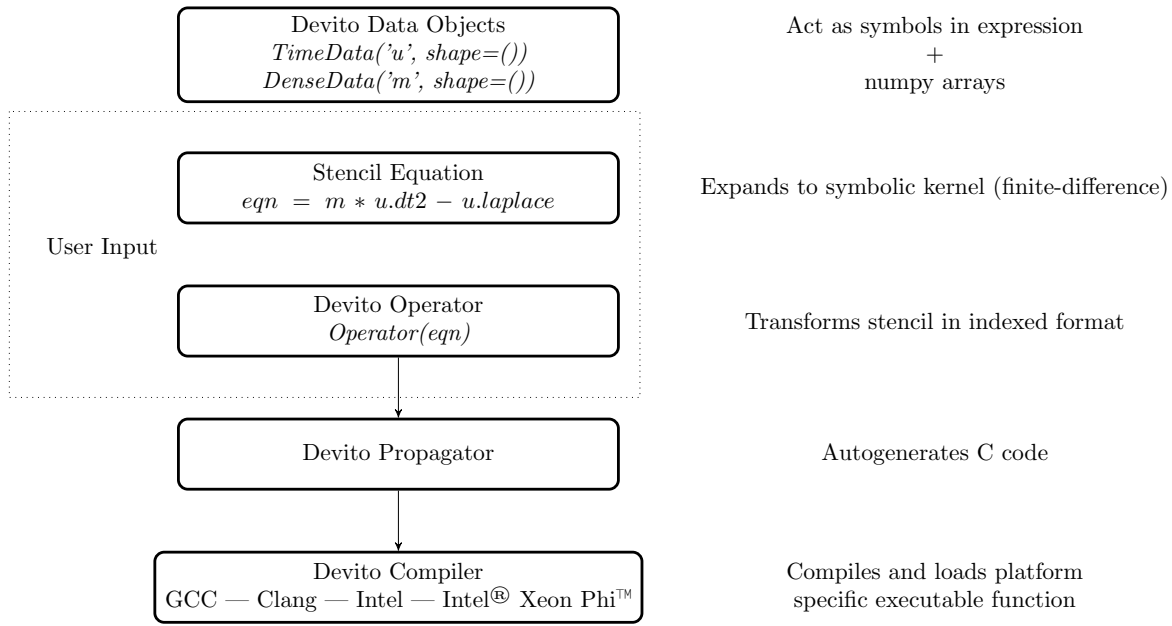

Figure 1: An overview of Devito's architecture

*3) Loop blocking:* Even boosted by vectorization and parallelization the code will be limited by its memory access. The next optimization would be to alleviate the memory traffic by reducing main memory access. By changing a single parameter, the framework generates code that uses loop blocking. Loop blocking increases cache reuse by harnessing data locality [13]. However blocking is only implemented in the spatial dimensions so far. An important factor that dictates the performance gain from blocking is the chosen block size. Apart from a user defined block size, Devito supports auto-tuning to automatically choose the block size that provides the most performance improvement. There is also the option of using a *best-guess* block size which uses the size of the stencil in the generated code along with the size of the cache of the target platform which is all determined at run-time.

*4) Memory mapping:* The biggest problem that inversion related code needs to usually address is checkpointing the results of wave propagation. This is because the variables that hold the wave-field(s) in memory quickly reach petabytes in size when a real problem is to be addressed. It therefore becomes imperative to save the results to disk since they are needed for a later step but are too big to sit in memory [22]. While other forms of checkpointing are under evaluation, Devito currently uses *mmap* to make the kernel automatically spill the arrays to disk when they exceed the available memory. Combined with a Solid-State Drive, which may be used exclusively for the purpose of storing the checkpoints, this approach can support very large problem sizes without any increase in complexity.

While saving the time history at each time step is important for the forward wave propagation step in an inversion workflow, its more common to not need to store all the history in memory. This is the case the gradient calculation in the inversion workflow. This is also preferable since this storage involving mmap above is quite expensive. In such a case where all timesteps do not need to be stored, common practice is to maintain only a certain number of timesteps in memory. This number depends on the discretization of the equation, i.e. how many steps backward in time does the stencil access. Devito allows one to switch between the two forms of time stepping by changing a single flag as can be seen in Listings 1 and 2.

*5) Common sub-expression elimination:* Several algebraic manipulation techniques can be used to reduce the computational complexity of a non-trivial expression. For example, suitable sequences of product expansions and factorizations may reduce the operation count or expose time-independent sub-expressions. At the moment, however, Devito only applies a basic technique, implemented using pre-existing *SymPy* operators: common sub-expression elimination (CSE). While it may be argued that CSE is already applied by most modern general-purpose compilers, there are three fundamental reasons that motivate the application of this technique at a higher level of abstraction:

**Readability of the generated code** Complex tensor equa-tions result in impenetrable scalar expressions in C.

**Compilation time** It can take more than a few hours to translate a moderately complex equation from C to machine code if CSE is not used and compiler optimization is switched to the maximum level.

**Fast Algebraic Transformation** An effective algebraic transformation system for complex finite difference expressions will consist of several steps. For example, it has been shown in [15] that transformations such as expansion, factorization, and code motion need to be composed in a clever way to achieve systematic reductions in operation count. Without CSE, applying these transformations would be prohibitively expensive in terms of code generation time.

## V. SEISMIC INVERSION EXAMPLE

Modern seismic exploration is heavily based on accurate solutions of wave equations in two cases: modelling and inversion. This requires agile software as the code will be executed millions of times over its life for different grid sizes, different model sizes and different time interval and sampling. Explicitly written code is, therefore, not the best approach as implementing loops and constants with inputs will drastically impact the efficiency of the solver as the compiler will be unaware of the grid sizes, for example. Case by case code generation is the the obvious solution to generate optimized code here. As modelling is part of the inversion, we will look directly at the complete picture.

### A. Problem definition

The simplest wave equation used is the acoustic case. For a spatially varying velocity model, $c$, the equation in the time domain is given by:

$$\begin{cases} m\frac{d^2u(x,t)}{dt^2} - \nabla^2 u(x,t) = q \\ u(.,0) = 0 \\ \frac{du(x,t)}{dt}|_{t=0} = 0 \end{cases} \quad (1)$$

where $u$ is the wavefield, $q$ is the source, $\nabla$ is the Laplacian and $\frac{\partial^2 u}{\partial t^2}$ is the second-order time derivative. This equation is solved explicitly with a time marching scheme. We are then looking at the following seismic inversion problem [9, 27] :

$$\underset{\vec{m}}{\text{minimize}}\,\Phi_s(\vec{m}) = \frac{1}{2}\left\|\mathbf{P}_r\mathbf{A}^{-1}(\vec{m})\vec{q}-\vec{d}\right\|_2^2, \quad (2)$$

where $\mathbf{A}$ is the discrete wave equation operator, $\vec{d}$ is the field measured data and $\mathbf{P}_r$ is the restriction operator to the measurement locations. To solve this optimization problem using a gradient-descent method we use the adjoint-state method to evaluate the gradient $\nabla\Phi_s(\vec{m})$ [18, 9]:

$$\nabla\Phi_s(\vec{m}) = \sum_{t=1}^{n_t} \vec{u}[t]\vec{v}_{tt}[t] = \mathbf{J}^T\delta\vec{d}, \quad (3)$$

where $\delta\vec{d} = \left(\mathbf{P}_r\vec{u}-\vec{d}\right)$ is the data residual, and $\vec{v}_{tt}$ is the second-order time derivative of the adjoint wave equation computed backwards in time:

$$\mathbf{A}^*(\vec{m})\vec{v} = \mathbf{P}_r^*\delta\vec{d}. \quad (4)$$

As we can see, the adjoint-state method requires a wave-equation solve for both the forward and adjoint wavefields and the full storage of the forward wavefield $\vec{u}$ in order to compute the gradient. While this computational cost clearly motivates the interest in optimizing the performance of the solvers, the importance of an accurate and consistent adjoint model in the solution of the optimization problem motivates the requirement to keep the implementation relatively simple.

### B. Testing framework

To ensure the accuracy of the generated code we have built a testing framework enforcing that our implementation satisfies the mathematical properties we expect from theory. The first mandatory property is to ensure that our derivation of the adjoint of the wave equation, obtained by taking the adjoint of a coded function, implements the adjoint of the operator. The mathematical test we use is

$$\text{for any random } \vec{x} \in \text{span}(\mathbf{A}^T),\ \vec{y} \in \text{span}(\mathbf{A}) \quad (5)$$
$$\langle \mathbf{A}\vec{x},\vec{y}\rangle - \langle \vec{x},\mathbf{A}^t\vec{y}\rangle = 0, \quad (6)$$

We then need to check that the gradient obtained by the code implementation of equation 3 satisfies the behaviour defined by the Taylor expansion of the objective function given in equation 2. Mathematically, this test is expressed by

$$\Phi_s(\vec{m}+h\vec{dm}) = \Phi_s(\vec{m}) + \mathcal{O}(h) \quad (7)$$
$$\Phi_s(\vec{m}+h\vec{dm}) = \Phi_s(\vec{m}) + h(\mathbf{J}[\vec{m}]^T\delta\vec{d})\vec{dm} + \mathcal{O}(h^2), \quad (8)$$

and is computed directly by varying the value of $h$ between $10^{-6}$ and $10^0$. By plotting the results on a loglog scale we can check that our implementation of the forward modelling behaves as a first order approximation and the gradient as a second order approximation.

### C. Implementation in Devito

Listing 1 shows the code required to solve the wave equation in a forward propagation run. The implementation first creates the required symbolic data object, spatial grid variables m and damp to represent the square slowness and spatially varying absorption coefficients respectively, and the time-dependent variable u to represent the pressure wavefield. The symbolic objects are then used to define the wave equation and the *SymPy* solve function is used to symbolically reorganize the resulting stencil. When creating the Operator instance the grid spacing and timestep size are inserted via a substitution dictionary subs, and the number of timesteps is defined before the operator is applied to the data. It is important to note here that the spatial order is defined by a single input parameter and thus easily changeable, and that the dimension of the problem setup is entirely defined by the shape of the input data array `model`.

Listing 1: Example code to solve the wave equation
```
def forward(model, nt, dt, h,
   spc_order=2):
  m = DenseData("m", model.shape)
  m.data[:] = model
  u = TimeData(name='u',
      shape=model.shape, time_dim=nt,
      time_order=2,
      space_order=spc_order, save=True)
  damp = DenseData("damp", model.shape)

  # Derive stencil from symbolic equation
  eqn = m * u.dt2 - u.laplace + damp *
      u.dt
  stencil = solve(eqn, u.forward)[0]

  # Add spacing substitutions
  subs = {s: dt, h: h}
  op = Operator(stencils=Eq(u.forward,
      stencil), nt=nt,
      shape=model.shape, subs=subs)
  op.apply()
```

The corresponding adjoint operator for the wave equation is demonstrated in Listing 2. The setup follows the same principles as the forward model except for subtracting the dampening term and using `u.backward` to define the symbolic stencil reorganization. The shorthand notation `u.forward` and `u.backward` hereby denote the highest and lowest stencil point in the second-order time discretization stencil, `t + s` and `t - s` respectively. This ensures the alignment of the final grid indexes after the spacing variables `s` and `h` have been resolved to explicit grid accesses in the final stencil expression.

Listing 2: Example code to solve the adjoint of the wave equation
```
def adjoint(model, nt, dt, h,
   spc_order=2):
  m = DenseData("m", model.shape)
  m.data[:] = model
  v = TimeData(name='v',
      shape=model.shape, time_dim=nt,
      time_order=2,
      space_order=spc_order, save=True)
  damp = DenseData("damp", model.shape)

  # Derive stencil from symbolic equation
  eqn = m * v.dt2 - v.laplace - damp *
      v.dt
  stencil = solve(eqn, v.backward)[0]

  # Add spacing substitutions
  subs = {s: dt, h: h}
  op = Operator(stencils=Eq(u.backward,
      stencil), nt=nt,
      shape=model.shape, subs=subs,
      forward=False)
  op.apply()
```

Listing 3 shows the C code generated by Devito when this function is called with $nt = 100$, $dt = 0.01$, $h = 0.1$ on an initial model of shape $(100, 100)$. A 2D model was chosen here for brevity of code, however changing the shape of the initial model to $(100, 100, 100)$ would be the only thing required to generate a 3D C code.

Listing 3: Generated C code from the forward operator
```
// #include directives omitted for brevity
extern "C" int ForwardOperator(double *u_vec, double *damp_vec, double *m_vec, double *src_vec, float
    *src_coords_vec, double *rec_vec, float *rec_coords_vec, long i1block, long i2block)
{
  double (*u)[130][130][130] = (double (*)[130][130][130]) u_vec;
  double (*damp)[130][130] = (double (*)[130][130]) damp_vec;
  double (*m)[130][130] = (double (*)[130][130]) m_vec;
  double (*src)[1] = (double (*)[1]) src_vec;
  float (*src_coords)[3] = (float (*)[3]) src_coords_vec;
  double (*rec)[101] = (double (*)[101]) rec_vec;
  float (*rec_coords)[3] = (float (*)[3]) rec_coords_vec;
  {
    #pragma omp parallel
    for (int i4 = 0; i4<149; i4+=1)
    {
      {
        #pragma omp for schedule(static)
        for (int i1b = 1; i1b<129 - (128 % i1block); i1b+=i1block)
          for (int i2b = 1; i2b<129 - (128 % i2block); i2b+=i2block)
            for (int i1 = i1b; i1<i1b+i1block; i1++)
              for (int i2 = i2b; i2<i2b+i2block; i2++)
              {
                #pragma omp simd aligned(damp, m, u:64)
                for (int i3 = 1; i3<129; i3++)
                {
                  double temp1 = damp[i1][i2][i3];
                  double temp2 = m[i1][i2][i3];
                  double temp4 = u[i4 - 1][i1][i2][i3];
                  double temp5 = u[i4 - 2][i1][i2][i3];
                  u[i4][i1][i2][i3] = ...
                }
              }
        for (int i1 = 129 - (128 % i1block); i1<129; i1++)
          for (int i2 = 1; i2<129 - (128 % i2block); i2++)
          {
            #pragma omp simd aligned(damp, m, u:64)
            for (int i3 = 1; i3<129; i3++)
            {
              double temp1 = damp[i1][i2][i3];
              double temp2 = m[i1][i2][i3];
              double temp4 = u[i4 - 1][i1][i2][i3];
              double temp5 = u[i4 - 2][i1][i2][i3];
              u[i4][i1][i2][i3] = ...
            }
          }
        for (int i1 = 1; i1<129; i1++)
          for (int i2 = 129 - (128 % i2block); i2<129; i2++)
          {
            #pragma omp simd aligned(damp, m, u:64)
            for (int i3 = 1; i3<129; i3++)
            {
              double temp1 = damp[i1][i2][i3];
              double temp2 = m[i1][i2][i3];
              double temp4 = u[i4 - 1][i1][i2][i3];
              double temp5 = u[i4 - 2][i1][i2][i3];
              u[i4][i1][i2][i3] = ...
            }
          }
        // Source and Receiver code omitted for brevity
      }
    }
  }
  return 0;
}
```

### D. Results

We illustrate our API on a 3D model. The model is 201 x 201 x 70 grid points with 40 absorbing boundary grid points on every side making the full computational size 281 x 281 x 150 grid points. The grid size is $15m$ and the source term is a Ricker wavelet at 10Hz. The wavefield is modelled for 1 second with 14th order spatial discretization. A snapshot of the wavefield is shown in Figure 2.

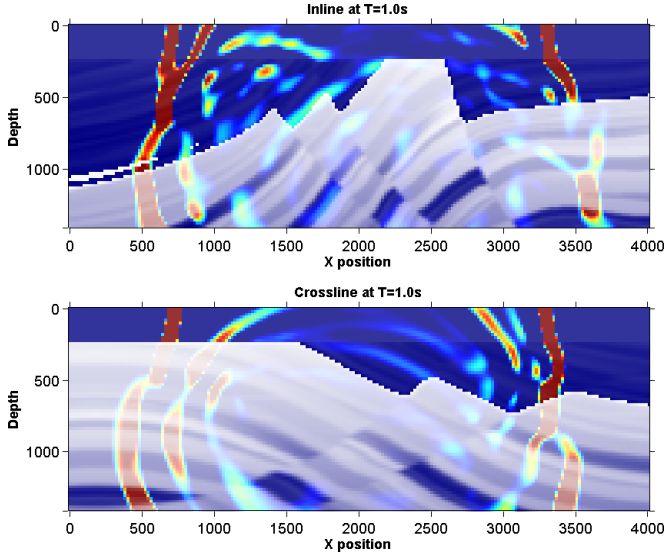

Figure 2: 3D acoustic modelling.

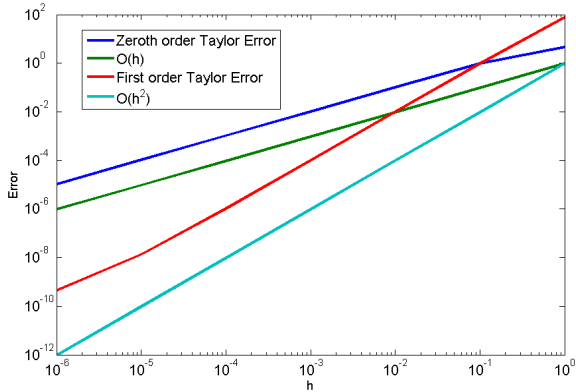

Figure 3: Gradient test for the acoustic kernel

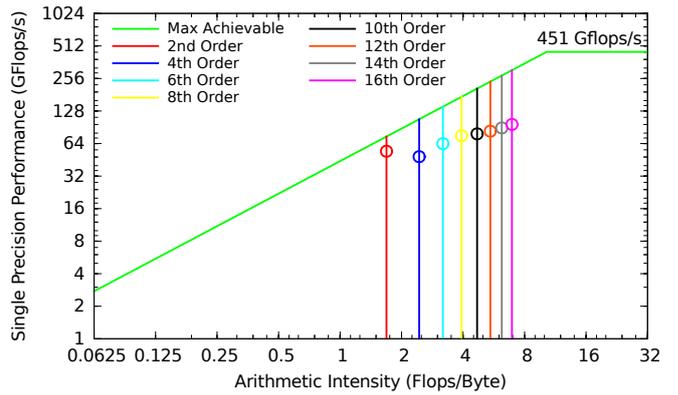

(a) Xeon

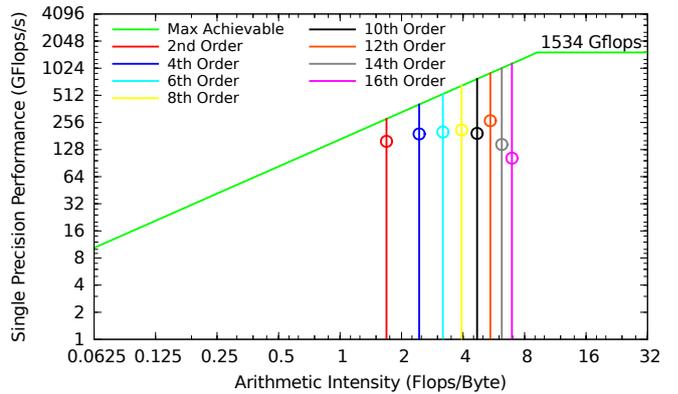

(b) Intel® Xeon Phi™

Figure 4: Roofline plots showing compute performance on two target systems

We validate our workflow by computing the error terms of Equation 7 for different value of $h$ and comparing it with the theoretical order.

Figure 3 shows that the gradients calculated by Devito satisfy the Taylor expansion properties and are therefore exact for the inversion problem, which guarantees the correctness of any gradient base method.

Figure 4 depicts the performance of 8 different versions of the generated code on an Intel® Xeon® E5-2690v2 10C 3GHz and a Intel® Xeon Phi™in comparison with the maximum achievable performance on each of the two platforms.

## VI. FUTURE WORK

We have two short-term goals for Devito: (i) the extension and formalization of the mathematical syntax used to express the stencils, and (ii) the improvement of the framework performance.

The mathematical language provided by Devito still lacks a proper formalization. At present, the *SymPy* 's language, enriched with a small set of Devito operators, is provided. New types of operators will be offered out-of-the-box (e.g., least square operators). Not only will these be essential for our driving application, seismic inversion, but also for a large class of problems arising in the most disparate fields.

We aim to exploit the mathematical layer to optimize the performance of the generated stencils, in particular to reduce their operation count through a suitable algebraic manipulation. Besides CSE, the fundamental transformations that we are considering at the moment are factorization and code motion (time-invariant sub-expressions may be exposed through factorization). Our ambition is to develop a system capable of optimizing generic stencils given a simple objective function (e.g., minimize the operation count without exceeding a certain working set size). A similar approach has been presented in [15]. Once the mathematical syntax is formalized and this expression rewriting engine is operative, domain specialists will not have to worry any more about the arithmetic complexity of the equations produced.

The current version of Devito implements simple systems for loop blocking and auto-tuning. We aim to replace these two components by integrating either a polyhedral compiler or a complete new abstraction, such as the already mentioned OPS. We are skeptic about the introduction of complex transformations such as time tiling: in a time loop iteration of full waveform inversion, for example, not only are multiple stencils applied, but also point evaluation over irregular grid regions needs to be performed. This clearly makes time blocking cumbersome. Nevertheless, these tools may improve vectorization and blocking over the spatial dimensions, as well as provide multi-backend support (e.g., GPUs).

## VII. Conclusions

In this article, we introduced Devito, a new framework for the expression of finite difference methods through a symbolic language. Its design is inspired and driven by real-world applications, in particular seismic inversion. Despite being in its infancy, Devito already provides a prototype DSL as well as a minimal set of features that make it possible to solve simple yet realistic models based on wave equations. The research presented in this work has a multidisciplinary nature, which is reflected by the multi-layer structure of the framework. This is perhaps one of the greatest strengths of the whole project. As explained in Section VI, our ambition is to create a tool which is actually usable by domain specialists. In fact, their feedback is already playing a fundamental role in the formalization and the development of the Devito DSL. Lastly, we remark that a necessary condition for the success of the project is the generation of highly optimized stencil code, as this is a strong requirement for the execution of real-world seismic inversion problems. A preliminary performance analysis has been presented in Section V-D.

## VIII. Acknowledgments

This work was financially supported in part by the Natural Sciences and Engineering Research Council of Canada Collaborative Research and Development Grant DNOISE II (CDRP J 375142-08), the Imperial College London Intel Parallel Computing Centre, British Gas, SENAI CIMATEC and the MCTI (Federal Govt of Brazil) scholarship MCTI/FINEP/CNPq 384214/2015-0. This research was carried out as part of the SINBAD II project with the support of the member organizations of the SINBAD Consortium.